\documentclass[conference]{IEEEtran}
\IEEEoverridecommandlockouts
\usepackage{cite}
\usepackage{amsmath,amssymb,amsfonts}
\usepackage{algorithmic}
\usepackage[ruled,vlined, linesnumbered]{algorithm2e}
\usepackage{graphicx}
\usepackage{textcomp}
\usepackage{tikz}
\usepackage{xcolor}
\def\BibTeX{{\rm B\kern-.05em{\sc i\kern-.025em b}\kern-.08em
    T\kern-.1667em\lower.7ex\hbox{E}\kern-.125emX}}
    
\usepackage{float}

\newcommand{\C}{\mathbb{C}}

\newcommand{\supp}{\operatorname{supp}}
\newcommand{\st}{\text{ subject to }}
\newcommand{\argmin}{\operatorname*{argmin}}
\newcommand{\argmax}{\operatorname*{argmax}}
\newcommand{\prox}{\operatorname{prox}}

\newcommand{\thres}{\operatorname{shrink}}
\newcommand{\Thres}{\operatorname{Shrink}}

\newcommand{\T}{\mathcal{T}}
\renewcommand{\L}{\mathcal{L}}

\usepackage{atbegshi}
\AtBeginDocument{\AtBeginShipoutNext{\AtBeginShipoutDiscard}}
    
\begin{document}

\title{Explicit CSI Feedback Compression via Learned Approximate Message Passing}

\thanks{Research is supported by a Nokia research donation. BG and GW are also supported by DFG Grant
with CoSIP ("Compressed Sensing in Information Processing" ) priority program Wu 798/8-2}

\author{\IEEEauthorblockN{1\textsuperscript{st} Benedikt Gro\ss}
\IEEEauthorblockA{\textit{Department of Computer Science} \\
\textit{Freie Universit\"at Berlin}\\
Berlin, Germany \\
benedikt.gross@fu-berlin.de}

\and
\IEEEauthorblockN{2\textsuperscript{nd} Rana Ahmed Salem}
\IEEEauthorblockA{\textit{Nokia Bell Labs} \\
Stuttgart, Germany \\
}
\and
\IEEEauthorblockN{3\textsuperscript{rd} Thorsten Wild}
\IEEEauthorblockA{\textit{Nokia Bell Labs} \\
Stuttgart, Germany \\
}
\and
\IEEEauthorblockN{4\textsuperscript{th} Gerhard Wunder}
\IEEEauthorblockA{\textit{Department of Computer Science} \\
\textit{Freie Universit\"at Berlin}\\
Berlin, Germany\\
}
}
\maketitle

\begin{abstract}
Explicit channel state information at the transmitter side is helpful to improve downlink precoding performance for multi-user MIMO systems. In order to reduce feedback signalling overhead, compression of Channel State Information (CSI) is essential. In this work different low complexity compressed sensing algorithms are compared in the context of an explicit CSI feedback scheme for 5G new radio. A neural network approach, based on learned approximate message passing for the computation of row-sparse solutions to matrix-valued compressed sensing problems is introduced. Due to extensive weight sharing, it shares the low memory footprint and fast evaluation of the forward pass with few iterations of a first order iterative algorithm. Furthermore it can be trained on purely synthetic data prior to deployment. Its performance in the explicit CSI feedback application is evaluated, and its key benefits in terms of computational complexity savings are discussed. 
\end{abstract}


\section{Introduction}

\subsection{Explicit CSI Feedback}
In 5G new radio, significant gains in data throughput and coverage are promised with an increase in number of antenna ports at gNB (basestation in new radio (NR)) and user equipment (UE) side, as well as advanced precoding schemes. These gains, however, are conditioned on accurate channel state information (CSI) at the gNB \cite{articleMarzetta}.  CSI at gNB side is used for multi-user grouping, downlink (DL) precoder design, frequency selective scheduling etc.
In time division duplex (TDD) systems, the channels in the uplink (UL) and DL directions are the same, hence gNB can deduce DL CSI from UL pilots. In frequency division duplex (FDD) systems, UE has to quantize and feedback the DL CSI to gNB in UL direction. In order to efficiently use UL resources, the UE has to compress the DL CSI before feedback to the gNB, hence trying to achieve a trade-off between UL resource utilization and CSI accuracy at gNB. Moreover, the compression step at UE must take into consideration the UE capability. For example, Type II Release 15 CSI codebook is considered as an optional UE capability since it requires high UE complexity.
In Release 17 MIMO agenda for CSI enhancements exploiting FR1 FDD reciprocity, the gNB has the task to learn the positions of the dominant taps from the explicit channel measured from UL sounding reference symbols (SRS) into a transform domain, which is then used to precode the pilots in the downlink (DL). A better system performance will naturally be observed with better compression schemes, since this will lead to a more accurate precoding of the pilots \cite{R1-2101011}. Hence, in Release 17 the problem of CSI compression is moved from the UE to the basestation. In this work, we assume the CSI compression to be taking place at the UE side, however same conclusions will naturally be valid when it is the task of the basestation to compress the CSI from the UL channel.

Compressed sensing (CS) methods have proven to be highly efficient in various communication applications, due to their ability to exploit structures inherent in the signals of interest and the availability of scalable, low complexity solvers \cite{schreck2014robust,wunder2015sparse,wunder2012nearly,schreck2009limited}. 
Time domain compression based CSI feedback schemes have been studied extensively in the literature. The main idea is to exploit the sparsity of the channel impulse response to achieve high compression. In \cite{tapLoc1,tap1_Q}, it was shown that by exploiting the sparsity of the channel impulse response, a
time domain analog CSI feedback scheme could outperform a frequency domain analog CSI feedback scheme. In \cite{ahmed2019comparison}, a time domain compression scheme was shown to outperform a frequency domain scheme based on principle component analysis. 
As explained in \cite{ahmed2018explicit}, owing to the presence of the guard bands in every OFDM symbol \cite{36873,38901}, the CSI knowledge at the basestation side is incomplete in frequency domain. In \cite{ahmed2018explicit,ahmed2019comparison}, compressive sensing (CS) algorithms were applied for sparse recovery of the time domain channel to discover the channel support, i.e., the location of the significant time domain channel taps. Therein the common channel support assumption was exploited, i.e. the location of the significant taps is the same for all transmit receive paths, which also agrees with the 3GPP channel model \cite{36873,38901}, in order to reduce the computational complexity of the CS scheme and also to achieve a noise reduction effect.

The 5G standard, currently being rolled out in many countries, is expected to benefit from the growing advancements in the AI/ML area to optimize some of its building blocks in the upcoming releases \cite{22.874}. The  future  generation  of  wireless  networks 6G is however expected to strongly capitalize on AI/ML enabled schemes as a main enabler in order to increase the system capabilities and at the same time reduce the complexity requirements at gNB and UE side \cite{6G_Deb}.

In this work we view this explicit CSI feedback in a two-stage process. In the first step, the CSI measurement at the UE is approximated by a structured sparse matrix in the delay domain using a compressed sensing algorithm. The second step then consists of selection of the strongest coefficients of this sparse approximation, followed by a quantization scheme. The location of the non-zero entries of the selected taps and their quantized values are then transmitted to the gNB. Assuming slow movement of the UE, the location of the taps is updated at a lower frequency than the transmitted values, and thus the computationally expensive step of finding a good sparse approximation via compressed sensing at the UE is only executed seldomly, thus saving further computational resources. Here, we focus on the first step, i.e. the sparse approximation and compare different low complexity algorithms in terms of reconstruction error in the frequency domain. We compare orthogonal matching pursuit (OMP) \cite{pati1993orthogonal, tropp2007signal}, a greedy heuristic that often performs well in practice, normalized iterative hard thresholding (NIHT) \cite{blumensath2010normalized}, which is based on hard thresholding and equipped with a linesearch strategy, the fast iterative soft thresholding algorithm (FISTA), an accelerated first order method based on convex relaxation, and introduce a neural network (NN) solution, termed learned approximate message passing for multi-measurement vectors (L-AMP-MMV). Its layers are tailored to the specific problem at hand, resulting in an efficient algorithm for CSI compression. Due to extensive weight sharing between the layers, the memory requirements for the neural network are comparable to the classical algorithms, even with many layers. 
In the problem of CSI acquisition, obtaining the ground truth is challenging and can be time consuming. In this work, we train the neural network on purely synthetic sparse data. This can enable an 'in-factory' training of UEs or basestations in which such neural networks are deployed. Further limited training in the field can also help customize the neural network to the deployed environment.

The paper is organized as follows: First the system model is specified. Then the algorithms to be compared are introduced, as well as the applied post-processing procedure, which is necessary to ensure that all algorithms give rise to the same overhead in the downstream computations, i.e. quantization and transmission of the compressed CSI. Numerical simulations on data from a 3GPP channel model are used to evaluate the candidate algorithms. Finally, the results are discussed.
\subsection{Notation}
Vectors and scalars are written as lowercase letters, while matrices are denoted by uppercase letters. For a matrix $X$, $X^T$ is its transpose and $X^H$ its conjugate transpose. $X_{:,i}$ is the $i$th column of a matrix, and $X_{i,:}$ its $i$th row. The $l_2$ norm of a vector is denoted by $\|x\|_2$, while  $\|X\|$ is the Frobenius norm of a matrix. The '$l_0$' semi-norm, which counts the number of non-zero entries of a vector is given by $|x|_0=|\{i : x_i\neq 0\}|$. Its convex relaxation, the $l_1$-norm is denoted $\|x\|_1$. Similarly, $\|X\|_{2,0}$ is the number of non-zero rows of a matrix, with the matrix norm 
$\|X\|_{2,1}=\sum\limits_i\|X_{i,:}\|_2$ being its convex relaxation. The support of a vector or matrix is denoted by $\supp(X)=\{(i,j) : x_{ij}\neq 0\}$.
The proximal operator \cite{moreau1962fonctions} associated with a function $f$, evaluated at point $v$ is defined as:
\begin{equation}
    \label{eq:prox}
    \prox_f(v) := \argmin\limits_x f(x) + \frac12 \|x-v\|^2.
\end{equation}

\subsection{System Model}
The channel impulse response (CIR) is assumed to be well approximated by a sparse vector, consisting of unknown complex amplitudes at a few unknown delays, i.e. for all $i=1,\ldots,P$ pairs of transmit/receive antennas, the CIR is approximately $x_i = \sum\limits_{k=1}^s c_{i,k}\delta_k(t_{i,k})$, where $c_{i,k}\in\C$ are the complex amplitudes and $t_{i,k}\in (0,1)$ are the time delays, normalized by the cyclic prefix length.
The channel is measured in the frequency domain at $M$ frequencies determined by the utilized hardware. Discretizing the delay domain, each CSI measurement can be written as $\C^M\ni y_i=Fx_i$, where $x_i\in\C^N$, $|x_i|_0=s$ and $F\in\C^{M\times N}$ is an oversampled DFT matrix, its entries corresponding to all possible frequency/delay pairs. Choosing a higher oversampling factor for the delays results in smaller mismatch between the true delay values $t_k$ and their discretized versions, reducing basis mismatch, at the cost of increasing the coherence of the system matrix $F$, which is detrimental to most compressed sensing algorithms. It is assumed that in the delay domain all beams share the same sparsity pattern, i.e. $\supp(x_i)=\supp(x_j)$ for all $i,j\in 1,2,\ldots,P$. This results in the problem of recovering a matrix $X=[x_1,\ldots x_P]\in\C^{N\times P}$ from measurements
\begin{equation}
\label{eq:measurements}
    Y = FX + \eta,
\end{equation}
where $Y\in\C^{M\times P}$ are the measurements in the frequency domain, $F\in\C^{M\times N}$ is a Fourier-type matrix and the unknown CSI in the delay domain is $X\in\C^{N\times P}$, which consists of $s$-sparse columns all sharing the same support. $\eta$ is complex-valued Gaussian noise. The recovery problem can thus be stated as
\begin{equation}
\label{eq:problem}
    \argmin\limits_{X\in\C^{N\times P}} \|Y-FX\|_2^2 \quad \st \|X\|_{2,0}\leq s
\end{equation}

\section{Algorithms}
\subsection{Classical Algorithms}
Several classical compressed sensing algorithms come into consideration for estimating the CSI $X$ from the measurements $Y$. Following \cite{ahmed2018explicit, ahmed2019comparison}, orthogonal matching pursuit (OMP) \cite{pati1993orthogonal, tropp2007signal} serves as a baseline. OMP is a greedy method that, one by one, picks the basis element from $F$ that has the highest inner product with the current residual $Y-FX^{(k)}$ and computes the new iterate $X^{(k+1)}$ as the least squares solution on the selected basis vectors. This procedure is repeated $s$ times to obtain solution consisting of $s$ non-zero rows (i.e. $s$ sparse columns with common support). It is depicted in algorithm \ref{alg:omp}. 
 \begin{algorithm}
 \label{alg:omp}
    \DontPrintSemicolon
    \SetKwInOut{Input}{input}
    \Input{Sparsity $s$, data $Y$, matrix $F$}
    $I^0=\{\}$ \;
    $R^0=Y$ \;
    \For{$k=1, 2, \ldots, s$}{
    \CommentSty{Compute column with highest inner product:} \;
    $i^* = \argmax\limits_{i} \sum\limits_j |F_{:,i}^HR^{k-1}_{:,j}|^2$ \;
    \If{$i^*\in I^{k-1}$ \KwSty{or} $\sum\limits_j |F_{:,i^*}^HR^{k-1}_{:,j}|^2<\epsilon$}{
    \KwSty{break} \;
    }
    $I^{k} = I^{k-1} \cup \{i^*\}$ \;
    $X^k = \left(F_{I^k}^HF_{I^k}\right)^{-1}F_{I^k}^HY$ \;
    \CommentSty{Update residual:} \;
    $R^{k} = Y-F_{I^k}X^k$ \;
    }
    \caption{OMP}
    \end{algorithm}
We also consider the normalised iterative hard thresholding (NIHT) algorithm \cite{blumensath2010normalized}, which combines gradient descend steps with hard thresholding and a tailored step-size selection scheme to solve \eqref{eq:problem}. Therefore, the operator $\mathcal{T}_s$ is defined. For a candidate solution $\bar{X}^{(k+1)}$ in step $k+1$, obtained by gradient descend, it computes in a first step the support via hard thresholding, i.e. the $s$ rows with largest $l_2$ norm are selected, and in a second step the least squares solution on the estimated support.
\begin{align}
    I_{k+1} &= \argmax\limits_{\substack{I\subset \{1,\ldots, N\}\\ |I|=s}} \sum\limits_{i\in I}\|\bar{X}^{(k+1)}_{i,:}\|_2 \\
    X^{(k+1)} & = \argmin\limits_{\substack{X\in\C^{N\times P} \\ \supp(X)\subset I_{k+1}}} \frac12\|Y-FX\|.
\end{align}
The algorithm is shown in alg. \ref{alg:niht}.
   \begin{algorithm}
   \label{alg:niht}
    \DontPrintSemicolon
    \SetKwInOut{Input}{input}
    \SetKwInOut{Initialization}{initialization}
    
    \Input{Sparsity $s$, data $Y$, matrix $F$, $0<c<1$}
    \Initialization{$X_1=0$, $I_1=\supp(\T_s(F^HY))$}
    \For{$k=1,2,\ldots$}{
    \CommentSty{Compute gradient:} \;
    $G_k = F^H(Y-FX_k)$ \;
    \CommentSty{Initial stepsize:} \;
    $\mu = \frac{\|G_{k|_{I_k}}\|^2}{\|F_{|_{I_k}}G_{k|_{I_k}}\|^2}$ \;
    $\tilde{X}_{k+1} = \T_s(X_k+\mu G_k)$ \;
    \If{ \KwSty{not} $\supp(\tilde{X}_{k+1}) = I_k$}{
    \CommentSty{Backtracking until stepsize is found} \;
    \While{$\mu > (1-c)\frac{\|\tilde{X}_{k+1}-X_k\|^2}{\|F(\tilde{X}_{k+1}-X_k)\|^2}$}{
    $\mu = \mu/2$ \;
    $\tilde{X}_{k+1} = \T_s(X_k+\mu G_k)$ \;
    }
    }
    $X_{k+1} = \tilde{X}_{k+1}, I_{k+1} = \supp(\tilde{X}_{k+1})$ \;
    }
    \caption{Normalised Iterative Hard Thresholding}
    \end{algorithm}
    
Another popular solver is fast iterative soft thresholding (FISTA) \cite{beck2009fast}, a first order method based on convex relaxation of \eqref{eq:problem}. For FISTA, the non-convex function $\|\cdot\|_{2,0}$ is first replaced by its convex relaxation $\|\cdot\|_{2,1}$. Then an unconstrained formulation is obtained by penalizing the constraint $\|X\|_{2,1}\leq s$, leading to the convex, unconstrained problem
\begin{equation}
    \label{eq:convex_unconstrained}
    \argmin\limits_{X\in\C^{N\times P}} \frac12\|Y-FX\|^2_2 + \lambda \|X\|_{2,1}
\end{equation}
with a scalar parameter $\lambda>0$ that balances the trade-off between data fit and sparsity. FISTA alternates between a gradient step on the data fit term $f(X) := \frac12\|Y-FX\|_2^2$ and a proximal point operation that serves as a replacement of a gradient descend step for the non-smooth part of the objective, $g(X) := \lambda\|X\|_{2,1}$. The proximal operator associated with the function $\lambda\|\cdot\|_{2,1}$ is row-wise shrinkage, which, for a matrix $X\in\C^{N\times P}$, shrinks each row norm by $\lambda$,
\begin{equation}
    \label{eq:prox_21}
    \left[\Thres(X, \lambda)\right]_{i,:} := \frac{\max(0, \|X_{i,:}\|_2-\lambda)}{\|X_{i,:}\|_2}X_{i,:} 
\end{equation}
for $i=1,2, \ldots, N$. The method is known to converge to a solution, if the stepsize in the gradient descend step is smaller than $1/\beta$, where $\beta$ is the Lipschitz constant of the gradient of the smooth part $f(X)$ of the objective. 
A Nesterov acceleration scheme \cite{nesterov1983method} is employed to achieve faster convergence. Furthermore, a restart condition \cite{o2015adaptive} is included into the algorithm that resets the stepsize if the update direction is no longer gradient-related. To overcome the problem of how to choose the parameter $\lambda$, we use continuation, i.e. starting with a large value, in order to encourage sparsity in the solution and then decreasing it geometrically to put more emphasis on data fit. The algorithm (without continuation and restart condition for simplicity) is depicted in alg. \ref{alg:fista}.
    
\begin{algorithm}
\label{alg:fista}
\DontPrintSemicolon
\SetKwInOut{Input}{input}
\SetKwInOut{Initialization}{initialization}
\Input{Lipschitz constant $\beta$, $f(X)=\frac12\|Y-FX\|^2$, $g(X)=\lambda\|X\|_{2,1}$}
\Initialization{Set $X_0$, $Z_0=X_0, t_0=1$}
\For{$k=0,1,\ldots$}{
$R_k=X_k-\frac{1}{\beta}\nabla f(X_k)$ \;
$X_{k+1}=\prox_{\frac{1}{\beta}g}(R_k)$ \;
$t_{k+1}=\frac{1+\sqrt{4t_k^2+1}}{2}$ \;
$\alpha_k=1+\frac{t_k-1}{t_{k+1}}$ \;
$Z_{k+1}=X_k+\alpha_k(X_{k+1}-X_k)$ \;
}
\caption{FISTA}
\end{algorithm}

\subsection{Learned AMP}
Approximate message passing (AMP) \cite{vila2013expectation} is another popular convex method for solving the 'standard' compressed sensing problem
\begin{equation*}
    \argmin\limits_x \frac12 \|y-Fx\|^2 + \lambda\|x\|_1,
\end{equation*}
for vector-valued $x\in\C^n$.
Its iterates are computed in a similar fashion as those of FISTA, but with a threshold parameter that is depending on the current iterate, and an additional term in the gradient step, the so-called 'Onsager correction'. 
\begin{algorithm}
\label{alg:amp}
\DontPrintSemicolon
\SetKwInOut{Input}{input}
\SetKwInOut{Initialization}{initialization}
\Input{data $y$, matrix $F$}
\Initialization{Set $v_{-1}=x_0=0$}
\For{$t=0,1,\ldots$}{
$b_t = \frac{1}{M}|x_t|_0$ \;
$v_t = y-Fx_t+b_tv_{t-1}$ \;
$\lambda_t = \frac{\alpha}{\sqrt{M}}\|v_t\|_2$ \;
$x_{t+1} = \thres\left(x_t+F^Hv_{t};\lambda_t\right)$ \;
}
\caption{AMP}
\end{algorithm}
While enjoying the same theoretical convergence rate as FISTA, in practice it often converges with roughly an order of magnitude fewer iterations. Its drawback is, that it is very sensitive to the sensing matrix. 
The algorithm is adapted to the matrix-valued problem \eqref{eq:problem} and unrolled for $T$ iterations to be used as a blueprint for a neural network as done in \cite{borgerding2017amp}, resulting in the proposed learned AMP for multi measurement vectors (L-AMP-MMV) architecture. 
Here we exploit the common channel support property in channels with limited bandwidth, i.e. the location of the significant
taps is the same for all spatial paths.
Therefore, we replace the soft thresholding function in line 5 of the algorithm by the thresholding operator associated with $\|\cdot\|_{2,1}$. 
The Onsager correction term in line 2 is replaced by $b_t=\frac1M |X|_{2,0}$. 

\begin{figure}
    \centering
    \resizebox{\columnwidth}{!}{
    \begin{tikzpicture}
\draw (0,0) node(xin) {$X_{t-1}$};
\draw (0,-1.5) node(vin) {$V_{t-1}$};
\draw (0,-5) node(yin) {$Y$};

\draw (2.5,0) node(plus0) [circle, text centered, draw] {$+$};
\draw[->] (xin) -- (plus0);
\draw (2.5,-1.5) node(bt) [rectangle, text centered, draw] {$B_t$};
\draw (1,-1.5) coordinate(v2b) {};
\draw[-] (vin) -- (v2b);

\draw[->] (v2b) -- (bt);
\draw[->] (bt) -- (plus0);

\draw (6,0) node(shrink) [rectangle, text centered, draw] {$Shrink(\cdot, \lambda_t)$};
\draw[->] (plus0) -- (shrink);
\draw (1, -2.5) coordinate(v2b2) {};
\draw (6,-2.5) node(lambda) [rectangle, text centered, draw] {$\frac{\alpha_t}{\sqrt{M}}\|\cdot\|$};
\draw[-] (v2b) -- (v2b2);
\draw[->] (v2b2) |- (lambda);
\draw[->] (lambda) -- node[right]{$\lambda_t$} (shrink);

\draw (8.5,0) coordinate(s2o) {};
\draw (8.5,-1.5) node(onsager) [rectangle, text centered, draw] {$\frac{\beta_t}{M}\|\cdot\|_{2,0}$};
\draw[-] (shrink) -- (s2o);
\draw[->] (s2o) -- (onsager);
\draw (8.5,-3.5) node(mult) [circle, text centered, draw] {$\cdot$};
\draw[->] (onsager) -- (mult);
\draw[->] (v2b2) |- (mult);
\draw (8.5,-5) node(plus1) [circle, text centered, draw] {$+$};
\draw[->] (mult) -- (plus1);
\draw[->] (yin) --(plus1);

\draw (10,0) coordinate(s2x) {};
\draw[-] (s2o) -- (s2x);
\draw (10, -3.5) node(f) [rectangle, text centered, draw] {$F$};
\draw[->] (s2x) -- (f);
\draw (10,-5) node(minus) [circle, text centered, draw] {$-$};
\draw[->] (plus1) -- (minus);
\draw[->] (f) -- (minus);

\draw (12,0) node(xout) {$X_t$};
\draw[->] (s2x) -- (xout);
\draw (12,-5) node(vout) {$V_t$};
\draw[->] (minus) -- (vout);

\end{tikzpicture}
}
    \caption{One layer of L-AMP-MMV with parameters $\alpha_t, \beta_t, B_t$}
    \label{fig:lamp_layer}
\end{figure}
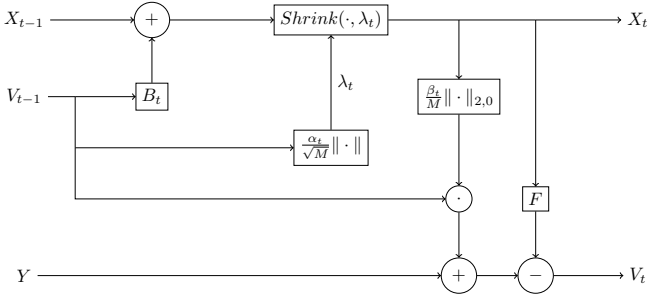

In order to turn this adaption of AMP to the row-sparse case into a neural network, we introduce some learnable parameters. In each layer, the matrix $F$ in alg. \ref{alg:amp} is replaced by a matrix $\alpha_t F$ with learnable scalar $\alpha_t$, and $F^H$ in line 5 of the algorithm by a learnable matrix $B_t$ that is initialized as $B_t=F^H$. Also, we allow the Onsager correction term in line 2 of alg. \ref{alg:amp} to be adjusted during training. Thus, the $t$th layer of L-AMP-MMV with parameters $\alpha_t, \beta_t, B_t$ receives inputs $(x_{t-1}, v_{t-1}, y)$ and computes
\begin{align}
    X_t &= \beta_t\Thres\left(X_{t-1}+B_tV_{t-1}; \frac{\alpha_t}{\sqrt{M}}\|V_{t-1}\|_2\right) \\
    V_t &= Y-FX_t + \frac{\beta_t}{M}\|X_t\|_{2,0}V_{t-1}
\end{align}
Here $\Thres$ refers to the soft thresholding operation defined in \eqref{eq:prox_21}.
A schematic of a L-AMP-MMV layer is depicted in fig. \ref{fig:lamp_layer}.
Following \cite{borgerding2017amp}, the parameters $B_t$ are shared across layers, $B_t=B$ for $t=0,1,.\ldots,T$ in order to significantly reduce the number of parameters of the model. The network is trained on synthetic sparse data, created according to \eqref{eq:measurements}, with the training schedule proposed in \cite{borgerding2017amp}, i.e. for $t=0,1,\ldots,T$ first only layer $t$ is trained for $n_{pre}$ epochs, while fixing the weights of all previous layers. After that all layers $1$ to $t$ are trained for $n_{post}$ epochs. The loss function for the training is
\begin{equation}
    \label{eq:loss}
    \L(\hat{X}, X, Y) = (1-\gamma)\|X-\hat{X}\|^2 + \gamma \|Y-F\hat{X}\|^2,
\end{equation}
where $0\leq \gamma\leq 1$ is a scalar parameter. Here, $X$ denotes a (synthetic) training sample, $Y$ the associated noisy input, and $\hat{X}$ is the output of the neural network.
The first summand in \eqref{eq:loss} drives the network towards computing sparse outputs, due to the training data $X$ being exactly row-sparse with sparsity at most $s$, while the second term aims at enforcing good data fit in the frequency domain. 

\subsection{Postprocessing}
The following post-processing strategy was applied to the computed solutions of the algorithms.
In the case of FISTA and L-AMP-MMV, the algorithm output was pruned to match its sparsity to the solutions of OMP and NIHT by setting to zero all but the $s$ rows with largest $l_2$ norm. Then the least squares solution was computed on the new support. This procedure enables a fair comparison of the solutions and ensures that the complexity in the downstream operations (i.e. quantization, transmission, etc) stays the same for all algorithms.

\section{Numerical Experiments}
 The algorithms were compared on test data from a 3GPP 3D dense urban macro (DUMa) channel model \cite{36873,38901} with 21 sectors. In each sector, 10 UEs were randomly dropped. At the gNB side, a 2x4x2 antenna array was used and the each UE is assumed to have 2 receive antennas. A first stage of spatial beamforming was assumed which resulted in an effective channel with 8 spatial beams for both polarizations. A subcarrier spacing of 15kHz was assumed and a carrier frequency of 2GHz. We assumed a channel frequency oversampling factor of 12, i.e., assuming one pilot subcarrier per physical resource block (PRB).
 The test set comprised 1050 measurements in the frequency domain, each consisting of 52 subcarriers measured over 16 spatial paths (8 beams x 2 receive antennas). The sensing matrix $F\in\C^{M\times N}$ has entries
 \begin{equation*}
 F_{m,n} = \frac{1}{\sqrt{1024}}\exp\left( \frac{-2\pi\cdot f_{m}n}{1024\cdot os} \right),
 \end{equation*}
 where $n = 0,1,\ldots, 256os$, $os$ is the oversampling factor, and the measured frequencies $f_m$ are chosen as every 12th subcarrier out of $[-312,-311,\ldots, 311]$. 
 It is noteworthy that this sensing matrix has bad RIP properties due to the regular subsampling of the rows of a DFT matrix. The data $Y$ was normalized to $\|Y\| = 1$.

\begin{figure}
     \centering
     \includegraphics[width=.45\textwidth]{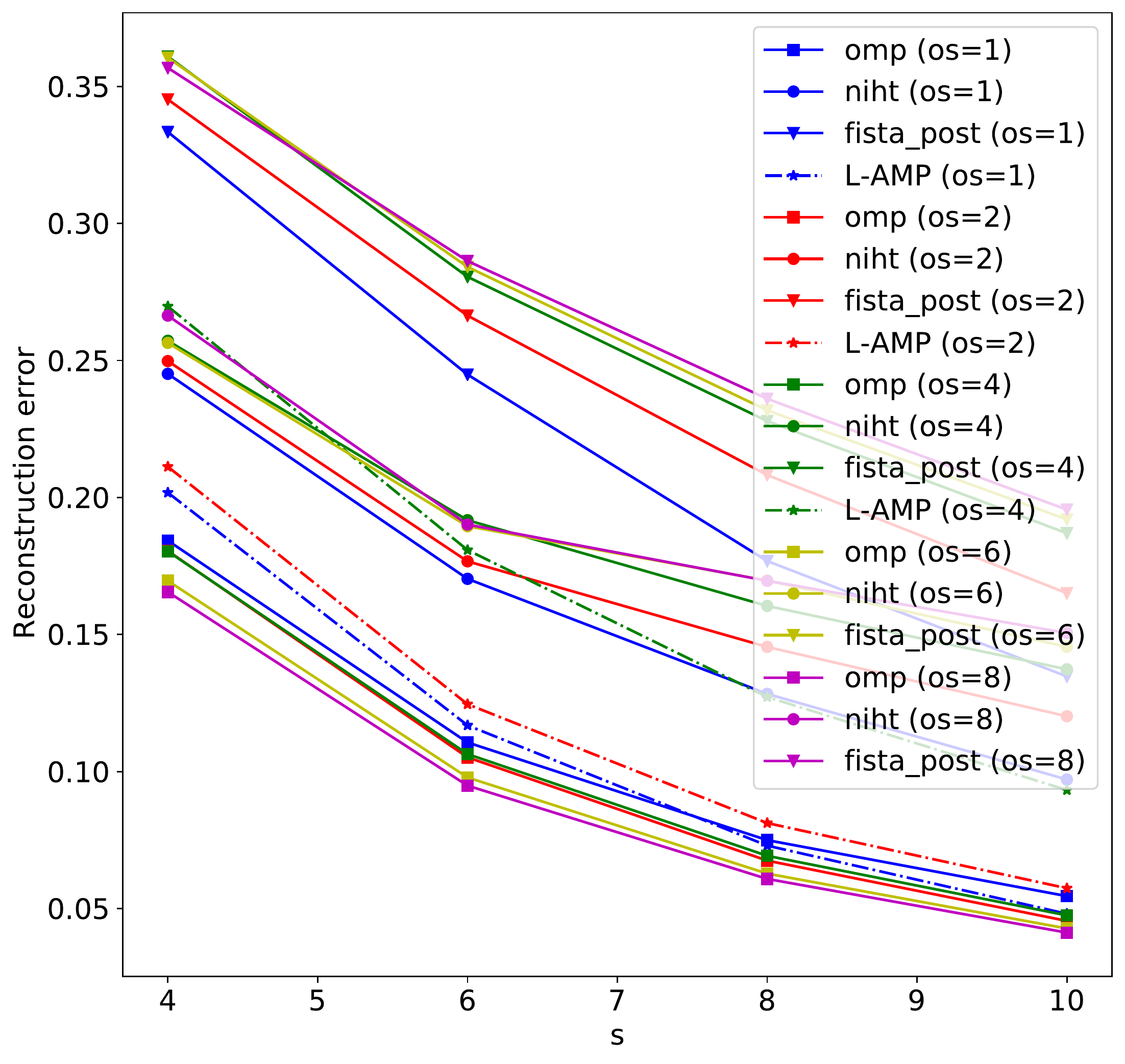}
     \caption{Mean reconstruction errors on the test set vs sparsity for different oversampling factors}
     \label{fig:classical}
 \end{figure}

 In figure \ref{fig:classical} the reconstruction error $\|Y-F\hat{X}\|_2$ between CSI data $Y$ and reconstruction $F\hat{X}$ for the classical CS algorithms OMP, NIHT and FISTA (+ postprocessing) are compared for various target sparsities $s$ and oversampling factors $os$. It can be seen that OMP not only outperforms the other approaches in terms of error magnitude, but is also the only algorithm that can take advantage of higher oversampling. NIHT and FISTA actually perform worse for higher oversampling, highlighting the fact that the sensing matrix does not posses suitable RIP properties.
 
 L-AMP-MMV was trained for a system with $os=1,2,4$ on synthetic data. Therefore, its parameters were initialized as $B=F^H$, $\alpha_t=\beta_t=1$ for all layers $t=1, \ldots, T$. The total number of layers was chosen to be $T=20$ in order to provide reasonable performance, thereby retaining low computational complexity during inference. Each layer was pretrained for 2 epochs, consisting of 1000 batches of 64 synthetically generated sparse samples. After each pretraining round, all layers up to the recently trained layer were fine-tuned for another 5 epochs.
 The training procedure is summarized in algorithm \ref{alg:training}.
 
 \begin{algorithm}
 \label{alg:training}
\DontPrintSemicolon
    \SetKwInOut{Input}{input}
    \SetKwInOut{Output}{output}
    \Input{System matrix $F$, number of layers $T$}
    Initialize $B_0=F^T$
    \For{$t=1,2,\ldots, T$}{
    Initialize layer $L_t$ with $\alpha_t=1, \beta_t=1, B=B_{t-1}$, ; \\
    Train layer $L_t$ for $n_{pre}$ epochs for fixed $L_1\ldots L_{t-1}$\\
    Train layers $L_1\ldots L_t$ for $n_{post}$ epochs.; \\
    }
    \Output{Trained model with Parameters $\left(\alpha_t, \beta_t \right)_{t=1}^T, B$}
\end{algorithm}

\begin{figure}
     \centering
     \includegraphics[width=.45\textwidth]{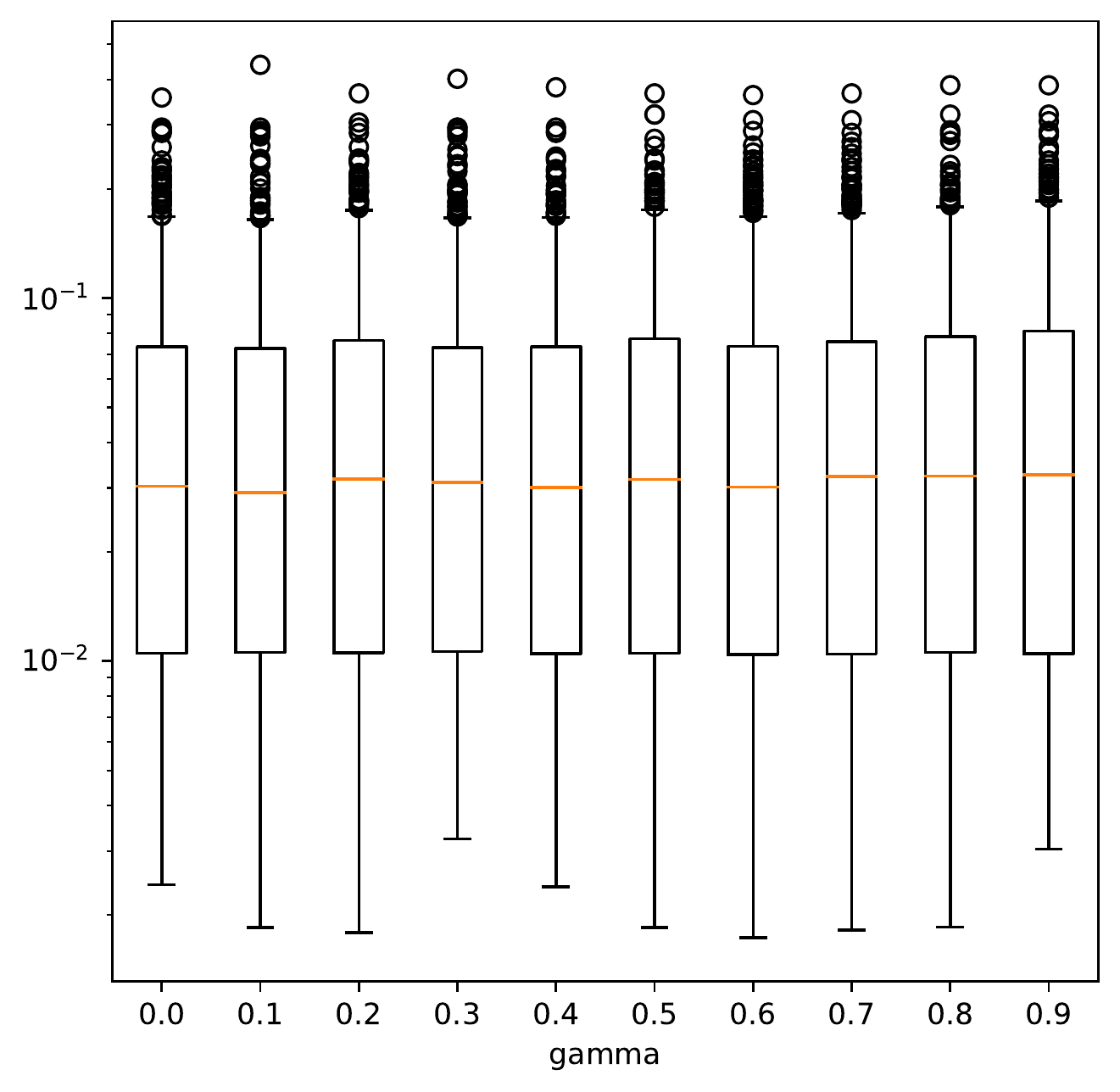}
     \caption{L-AMP reconstruction error distributions after post-processing for various $\gamma$ (os=4, s=10)}
     \label{fig:lamp}
 \end{figure}

 For the training procedure the ADAM optimizer \cite{kingma2014adam} with a learning rate of $10^{-3}$ was chosen. The data from the 3GPP channel model was used as test set.
 Figure \ref{fig:lamp} compares the reconstruction error of L-AMP-MMV (+ postprocessing) for several values of $\gamma$. $\gamma=1$ was excluded from the figure because its error was too high. It can be seen that, because of the applied postprocessing, the error is not very sensitive to the value of $\gamma$, only requiring $0\leq \gamma < 1$.
 The error in fig \ref{fig:classical} shows L-AMP-MMV is competitive to the best classical algorithm OMP. Since the number of parameters scales linearly in the oversampling factor, higher oversampling requires much longer training time in order to achieve the same or lower error. We chose to keep the number of training epochs constant to show this effect of oversampling on L-AMP. With longer training the reconstruction error is expected to go down further. Note that the time-consuming training of the L-AMP network is done offline on purely synthetic data prior to deployment. The trained network requires only a forward pass, consisting of $2T$ matrix multiplications and some thresholding to compress the CSI data. This is equivalent to the cost of $T$ iterations of an iterative first order algorithm like FISTA.

\section{Discussion and Conclusion}
\subsection{Discussion of the Results}
FISTA and NIHT do not reach the performance of OMP and L-AMP-MMV in terms of solution quality and computational complexity. While both algorithms have a very low per iteration complexity, basically only requiring 2 matrix multiplications and the soft/hard thresholding operation, the number of iterations needed to get satisfactory solutions (mostly above 100 in this simulations) causes them to exceed the computation needed for the baseline method OMP, which terminates after $s$ steps, each requiring the computation of the inner product of the residuals and the columns of the system matrix and a projection of the selected basis vectors. The neural network solution is very efficient in terms of complexity, since in each layer only 2 matrix multiplications, the efficient shrinkage operation and some norms have to be computed. 
The inferior performance of NIHT and FISTA is due to the fact that both algorithms rely on the system matrix $F$ satisfying a restricted isometry property (RIP) or low coherence, which is not the case for the matrix $F$ in question, especially with a high oversampling factor. The L-AMP-MMV neural network performs competitive with OMP but performs the computation faster, especially when implemented on a GPU, which takes advantage of the fact that NN operations are highly parallelizable. With training data that better matches the test data we expect the error to decrease further. In terms of computation, the neural network approach has significant advantages. Its operations are fully parallelizable and fit for implementation on GPUs or FPGAs, while OMP is sequential in nature. While training the neural network is costly, it can be done offline and, as the results have shown, on purely synthetic data. This is crucial for the application as compressor for explicit CSI feedback, since it is infeasible to measure large amounts of real CIRs of wireless communication channels. Furthermore, advances in synthetic data generation via generative models such as GANs and variational autoencoders offer a promising direction to improving the training process by creating more realistic data.

\subsection{Conclusion}
We have introduced a neural network design, L-AMP-MMV, for explicit CSI feedback compression of radio channel for next releases of 5G NR and for future generation of wireless networks 6G. While its performance in terms of approximation error was competitive to, but not better than the baseline method OMP, this approach offers attractive advantages for high throughput scenarios inasmuch it is highly parallelizable, low complexity and requires only little storage for the weights due to extensive parameter sharing across layers. The L-AMP-MMV was trained on purely synthetic data with no prior knowledge of channel model, which can be very advantageous for the CSI acquisition problem.
Improving the accuracy of L-AMP-MMV by employing more sophisticated methods for the generation of training data, as well as incorporating the subsequent quantization into the network design will be a topic of future research.

\bibliographystyle{unsrt}
\nocite{*}
\bibliography{refs}

\end{document}